\documentclass[12pt]{article}

 \usepackage{amsfonts}
 \usepackage{amssymb,amsmath}
 \usepackage{graphicx} \usepackage{epstopdf}
 \newcommand{\crlb}[1]{\label{#1}\\[2pt]}
 
 \newcommand{\crld}[1]{\label{#1}}
 \newcommand{\eela}[1]{\quad\hbox{\scriptsize{#1}}\label{#1}\end{eqnarray}}
 \newcommand{\eelb}[1]{\label{#1}\end{eqnarray}}
 
 \newcommand{\newsecb}[2]{\section{#1}\label{#2}\setcounter{equation}{0}}

 \newcommand{\nolabels} {\def\eel{\eelb}\def\eeql{\eeqlb}  \def\crl{\crlb} \def\newsecl{\newsecb}\def\bibiteml{\bibitem} \def\citel{\cite}\def\labell{\crld}}
\newcommand{\eeqla}[1]{\quad\hbox{\scriptsize{#1}}\label{#1}\end{aligned}\end{equation}}
\newcommand{\eeqlb}[1]{\label{#1}\end{aligned}\end{equation}}

\newcommand\publishversion{\nolabels\setlength{\textheight}{8.3in}\setlength{\oddsidemargin}{0in}
   	 \setlength{\textwidth}{6.3in}\setlength{\topmargin}{-0.2in}}

\def\beq{\begin{equation}\begin{aligned}}		\def\eeq{\end{aligned}\end{equation}}
\def\be{\begin{eqnarray}}  					\def\ee{\end{eqnarray}}		
\def\bi#1{\begin{itemize}\item[#1]} 			\def\itm#1{\item[#1]} 			\def\ei{\end{itemize}} 
  \def\eqn#1{(\ref{#1})}
   	 \def\fn{\footnote}	  		 

		       \def\d{\delta}   \def\k{\kappa}

	    		        		     		\def\vv{\varphi}     
 	 		     	\def\tht{\theta}      	 
	     		          		              	  
\def\W{\Omega}    		  		 		  
\def\OO{{\mathcal O}}

\def\ffract#1#2{\raise .2 em\hbox{$\scriptstyle#1$}\kern-.3em/\kern-.2em\lower .15 em \hbox{$\scriptstyle#2$}}
					
\def\tl#1{\tilde{#1}} 
 		\def\qqquad{\qquad\qquad}		
\def\bpmatrix{\begin{pmatrix}} 			\def\epmatrix{\end{pmatrix}}
\def\bmatrix{\begin{matrix}} 			\def\ematrix{\end{matrix}} 
\def\bcenter{\begin{center}}			\def\ecenter{\end{center}}

\def\lowerheightfig#1#2#3{\(\raise-#1\hbox{\includegraphics[height=#2]{#3}}\)}
\def\lowerwidthfig#1#2#3{\(\raise-#1\hbox{\includegraphics[width=#2]{#3}}\)}
		 
\def\weglaten#1{}
\def\BH{{\mathrm{BH}}}\def\Pl{{\mathrm{Pl}}}
\publishversion

\begin{document}

\begin{titlepage}
 \title{ \LARGE\bf What happens in a black hole
 \\[5pt] when a particle meets its antipode\\[10pt]}
\author{Gerard 't~Hooft}
\date{\small  Institute for Theoretical Physics \\ Utrecht University  \\[10pt]
 Postbox 80.089 \\ 3508 TB Utrecht, the Netherlands  \\[10pt]
e-mail:  g.thooft@uu.nl \\ internet: 
http://www.staff.science.uu.nl/\~{}hooft101/}
 \maketitle

\noindent \textbf{Abstract}\\ [5pt]
The notion of antipodal identification on the black hole horizon is further explained and elaborated. 
Contrasting with numerous attempts in the literature to make fuzzy, poorly motivated models for black holes, we explain how, with an absolute minimum of assumptions, known laws of local physics suffice to calculate the unitary evolution law for a Schwarzschild black hole. Earlier work by the author, which explains how firewalls can and must be avoided, while also the information paradox disappears, ran into
its one remaining problem: how to explain it better to the community. Antipodal identification is a natural way to replace  thermally mixed states by pure quantum states, without the need to hide our ignorance in ``chaos". We do encounter a strange looking sign switch in the relation of particles to their antipodes near the horizon. This sign switch is necessary to recover complete unitarity without any loss of information anywhere, while restoring locality. \\[5pt]
Although there are some important remaining problems, we advertise our approach as a healthy alternative to the reliance on AdS/CFT conjectures, which, we claim, do not guarantee to provide reliable answers.
 \end{titlepage}
 
\newsecl{Introduction. The antipodal identification}{intro} \setcounter{page}{2}
When a new `no-go theorem' appears in the physics literature, it is usually concluded that `new and different' physics is needed to resolve some paradox. However, history has told us that a more appropriate response may also be: 
`they overlooked something'. Much more strongly than often realised, such theorems hinge on assumptions written in small print. This author has seen several `no-go' theorems go up in flames. The so-called `black hole information paradox'\,\cite{SWH}---\cite{Giddings1} is a good example. As what happened to other paradoxes and no-go theorems, also this paradox was first attacked by outsiders who did not properly understand the details of the argument. The paradox is actually not easy to disentangle. The problem was well phrased by  Almheiri et al\,\cite{firewall,Polch}. It was claimed that, in a black hole,  at least one of the four principal requirements of a physical theory must be strongly violated. The requirements may be formulated as: 
\bi{1.} The process of formation and evaporation of a black hole, as viewed by a distant observer, can be described entirely within the context of standard quantum theory. In particular, there exists a unitary \(S\)-matrix that describes the evolution from ingoing matter to outgoing Hawking-like radiation. 
\itm{2.} Outside the stretched horizon of a massive black hole, physics can be described to good approximation by a set of quantised field equations in a semi-classical expansion. All quantum gravity effects die off rapidly as the distances from the event horizon increase.
\itm{3.} To a distant observer, a black hole appears to be an ordinary quantum system with distinct quantum states. The dimension of the subspace of states describing a black hole of mass \(M\) is the exponential of the Bekenstein entropy \(S(M)\).\fn{One author added: ``The vacuum of the theory is unique". This is true, but only in the following sense: the vacuum state is the state with lowest energy, where the black hole is absent. When the distant observer annihilates all Hawking particles, there will be no black hole left.}
\itm{4.}  A freely falling observer experiences nothing out of the ordinary when crossing the horizon.
\ei
These authors conclude that one cannot avoid the emergence of `firewalls', heavy showers of particles close to the future event horizon; other researchers opt for a fundamental form of non-local physics near the horizon\,\cite{Giddings2}---\cite{Giddings5}. The proponents of the AdS/CFT approach just say that the miracles of their mapping preclude a description of what goes on in the black hole in terms of local dynamical variables, and that the micro-states of a black hole simply tend to become `chaotic'. Only in the CFT branch they claim to understand what is going on. \def\max{{\mathrm{max}}}

Since the present author strongly suspects that in these papers the gravitational back reaction between in- and out going particles is not taken into account properly, he decided to do his own calculation of a unitary black hole evolution operator, obeying exactly the above four requirements, but based on an approach developed in the 1980s\,\cite{GtHstrings1} and 1990s\,\cite{GtHstrings2,GtHSmatrix}. 

The calculation is done in the limit \(M_\BH\gg 1\) in Planck units. In Schwarzschild coordinates, at space-time points separated further from the horizons than a few times the Planck length, we assume the validity of some ``Standard Model", with gravity added perturbatively in the form of gravitons -- exactly as in postulate \# 2. Every particle is assumed to have transverse momenta \(\tl p_i\) and intrinsic masses \(m_i\) all below a maximum \(m_\max\) obeying, in Planck units,
\be 1/M_\BH \ll m_\max \ll 1\ , \eel{mmax}
so that they do not bring about curvatures in the transverse direction  bigger  than \ 
\( \OO(1/m_\max)\ \) in Planck units. The angular momenta for individual particles near the horizon will stay below \(\ell_\max\), with 
\be 1\ll\ell_\max\ll M_\BH\ . \eel{ellmax}
This is good enough for the Hawking particles, which have energies comparable with \(1/M_\BH\), so that their angular momenta are of order 1. Also, the total number of particles needed to be considered in the vicinity of the black hole, as seen by a local observer, yet outside the stretched horizon, and within a time slot of length \(M_\BH\,\log M_\BH\) in Planck units, is taken to be of order 1. Consequently, the effects all these particles have on the Schwarzschild metric, may be considered to be negligible. 

We start with a bunch of in-going and out-going particles \(i=1,\cdots,N\) \emph{on} the horizon, whose transverse coordinates are \(\W=(\tht,\,\vv)\). In light cone coordinates, on either the future or the past event horizon, their longitudinal momentum operators are \(p_i^\pm\) and their longitudinal position operators are \(u^\pm_i\). Both horizons can function as (parts of) Cauchy surfaces. \(p^-_i\) and \(u^+_i\) and \(\W_i\), obeying \( [u^+_i,\,p^-_j]=i\d_{ij}\),  are the Cauchy data on the future horizon, and \(p^+_i\) ,  \( u^-_i\) and \(\W'_i\), with \([u^-_i,\,p^+_j]=\d_{ij}\),  are the Cauchy data on the past horizon. Ordinary physics (QFT on a curved metric background) determines how particles evolve from past to future horizons (partly also escaping to the asymptotic domain at infinity). `New' calculations are needed that explain how the black hole responds; what we are after is a unitary operator that translates the data entering at the future horizon, back to the past horizon where it re-emerges.\fn{This sounds like a violation of causality, but it is not. A time lapse will occur in the order of \(M_\BH\log M_\BH\) in Planck units.}  We define a momentum distribution \(p^\pm(\W)\) and a position distribution \(u^\pm(\W)\), such that, in longitudinal light cone coordinates,
\be p^\pm(\W)=\sum_ip^\pm_i\d^2(\W,\W_i)\qquad\hbox{and}\qquad u^\pm(\W_i)=u^\pm_i\ , \eel{momposdistr}
after which we define the spherical harmonic expansion,
\be p^\pm(\W)\equiv \sum_{\ell,\,m}p^\pm_{\ell,m}\,Y_{\ell,m}(\W)\ ;\qquad u^\pm(\W)\equiv\sum_{\ell,\,m}u^\pm_{\ell,m}\,
Y_{\ell,m}(\W)\ , \eel{sphericalharmonics}
obeying
\be{}[u_{\ell,m}^\pm\,,\ p_{\ell',m'}^\mp]=i\d_{\ell\,\ell'}\d_{m\,m'}\ . \eel{commrels}
The effects of the gravitational shifts are then written down in terms of these distributions.

Somewhat to his own surprise, the author found this calculation to be easy; we only use local, standard physics outside the horizon. Hoping to spark a discussion, this result was quickly put on the web\,\cite{GtHdiagonal}, but no discussion came. 

In most physical systems the \emph{information flow} is quite clear and needs no further explanations, but in most public discussions of quantum black holes, the implicit assumptions are very dubious. As soon as a classical or quantum particle enters the trapped region, only a miracle can bring it back. The classical geodesics all move away from that horizon. Very importantly, in our work, we do not require such miracles.

In contrast, what we do have is the gravitational back reaction. It so happens that this force displaces the information on both horizons. At first sight, this only exacerbates the problem, but upon closer inspection, we can indeed take this force into account, albeit not in the way usually attempted. The complete gravitational interaction is of course difficult to handle, but one can make an approximation that works well in the limit \(M_\BH\gg M_\Pl\). In this case, it is reasonable to expect that all particles under consideration, with masses \(m\) and transverse momenta \(\tl p\), have \(\tl p^2+m^2=\OO(M_\Pl^4/M_\BH^2)\ll M_\Pl^2\). We can then assume that, once close to an horizon, the particles will almost be moving with the speed of light, so that we may assume gravity to generate the gravitational equivalence of {C}erenkow radiation, causing test objects to be displaced by a finite amount when the particles pass by, in a surrounding space-time that, in all other respects, stays Minkowskian.

This approximation is only valid in the immediate neighbourhood of the horizon, but this is also the only place where we need this information. Most importantly, we can continue the calculations without being disturbed by additional curvature in space-time, while the displacement vector obeys simple equations.  

A problem will be that the only source of information comes from the distribution of the in-going momenta, and its only effects come in the form of displacements, while the beauty of this situation is that momenta and displacements are each other's dual conjugate in quantum mechanics.

Since we did the calculation as accurately as we could, the shortcomings of our thinking~-- everybody's thinking~-- at the time were clearly exposed. Indeed, closer inspection of our paper revealed what seemed to be a mistake: the evolution operator we derived there was not unitary, or more precisely, it was unitary in a hypothetical world containing \emph{two} identical black holes instead of one. Now this configuration has been studied by Maldacena and Susskind\,\cite{ER=EPR}, who interpret this as a description of `two entangled black holes', relying on the notion that when two distant objects are quantum mechanically entangled, they do not really exchange information, and unitarity will not be violated by any one of them. The idea that the `Einstein-Rosen bridge', connecting the metrics of these two black holes, relates to the entanglement seen in the `Einstein-Podolsky-Rosen paradox', was aptly nicknamed `ER=EPR'.

But this idea is also easy to refute. Regions \(I\) and \(II\) are entangled in the conventional sense, \emph{only if} we switch off the gravitational back reaction. Exactly the back reaction causes information to be passed on from region \(I\) to region \(II\) and back in the Penrose diagram, as is clearly shown in our explicit calculation of the evolution matrix. It enables one to  construct operators pertaining to the space-time region where one black hole occurs, and operators near the other black hole, whose commutators do not vanish, see Ref.\,\cite{GtHrecent1}, Appendix B. \  In QFT, this is a signal for a disastrous form of non-locality. A better cure for this problem had to be found.

Naturally, one may suggest that regions \(I\) and \(II\), somehow, represent \emph{the same} black hole. This however, would also lead to problems:  the gravitational back reaction for a particle in region \(I\), would exactly cancel the effect of the same particle in region \(II\). (see the discussion of this in Section~\ref{meeting}). Thus, the gravitational back reaction would cancel out everywhere, and we would end up with no information exchange at all. Again, this does not lead to a unitary evolution matrix.

But could region \(II\) represent the same black hole at some other value of the transverse angles \(\tht\) and \(\vv\)? The answer is yes, that could be possible, but we must not allow for the existence of  any fixed value of \(\tht\) and \(\vv\) for which region \(I\) does coincide with region \(II\) -- at such a point, the back reaction would vanish, which one should never allow. The condition that there be no such fixed points, leads to a constraint condition that allows only one solution: 

\qqquad \emph{Region \(II\) represents the antipodes of region \(I\). }

\noindent This we call the `antipodal identification', motivated and explained in more detail in Ref.\,\cite{GtHrecent1}, Appendix~A. \ It implies that, on the horizon,  \(r=2GM\), a point given by the  transverse coordinates \(\W\equiv(\tht,\,\vv)\) represents the same space-time point as the one with transverse coordinates \(-\W\equiv(\pi-\tht,\,\vv+\pi)\). Or, a curve that crosses the horizon from region \(I\) into region \(II\), does not enter anything like the `interior' of the black hole, but instead, goes directly to the points outside the antipode. 

Thus, the horizon changed from a 2-sphere into a projective 2-sphere. The antipodal identification on the horizon ensures that the entire metric has the same asymptotic structure at infinity as Minkowski space (without doublings). Instead of a mysterious `interior region', when crossing the intersection of the future and the past event horizon, we find ourselves outside the horizon again. We are in region \(II\), but now this means we arrived at the other side of the black hole.

At the absolute future\fn{The words `absolute future' here are used to indicate that all points in this sector are at the future of all points of the outside world, provided one follows an oriented geodesic within the local light cones; similarly, the `absolute past' is defined.} and the absolute past\fn{Also in the cited references, the author motivates his use of the \emph{eternal Penrose diagram}, which means that matter imploding in the distant past, as well as matter involved in the evaporation of the hole in de distant future, are here not taken into account; the diagram is chosen to describe the quantum behaviour in a relatively short time interval only. How to link different time intervals together can be investigated and explained later.} we have the trapped regions \(III\) and \(IV\). After a time reversal, these regions \(III\) and \(IV\) may also be seen as each other's antipodes.

No singularity has been added to the physically accessible regions. This is because the transverse component of the metric there, never gets a radius less than \(r=2GM_\BH\). A new singularity would be added at \(r=0\), but the Schwarzschild solution already had a singularity there, and, very importantly, this singularity is hidden far within the trapped regions, and later we shall explain why the trapped region, from any finite distance off the horizon onwards, does not play any role in the evolution operator.

Our work had been inspired, among others, by the results of Hawking, Perry and Strominger on the BMS group\,\cite{HPS}. These ideas may possibly be related to our methods. The physical interpretation of the `conserved charges' they report about is somewhat unclear to this author. These charges do seem to carry some resemblance to the ghost states associated to gauge fixing, as encountered in particle theory approaches; indeed, in our work it is important that the coordinates are chosen such that asymptotic space-time is manifestly flat. For this, `soft gravitons' (gauge transformation generators) will be of importance, but in our formalism they were avoided; by allowing the shifts in the coordinates (the apparently non-local gravitational drag) we adjust to `safe' coordinates in the asymptotic domains. In Ref.\,\cite{HPS}, also hints are given towards antipodal correlations.

\newsecl{The sign switch between regions \(I\) and \(II\)}{signs}
With the antipodal identification in place, our approach  appears to provide for a unitary evolution matrix. Information is passed on from the horizon directly to its antipodes. From a physical point of view, this seems to be quite acceptable. Indeed, in the terminology normally used in the old form of the metric, this implies a kind of non-locality as was foreseen by Giddings\,\cite{Giddings3}--\cite{Giddings5}. It is in our new, topologically non-trivial space-time, that all interactions are local again.

One may look at this the following way: a non-local interaction that would solve the problems observed by Almheiri et al\,\cite{firewall,Polch}, and Giddings, may be limited to a direct interaction between data at one point of the horizon and its antipode. What we propose is an extra topological twist in space-time that brings these points close together, so, in this space-time, the interaction stays local.
 
 \begin{figure}[h!]
	\qqquad \includegraphics[width=350pt]{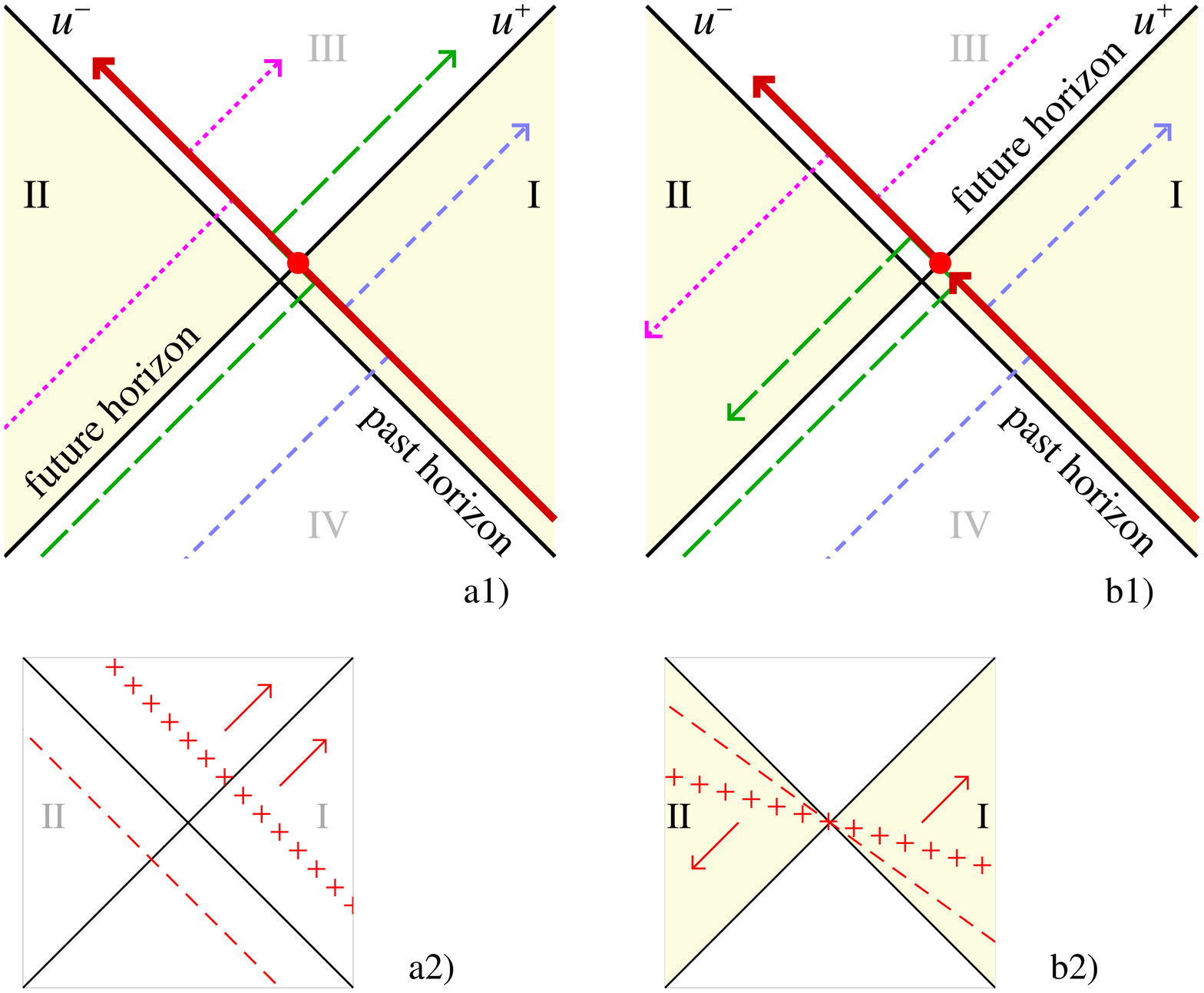} \begin{quote}
	\begin{caption}{ }{\small{$a1)$ As seen by a local observer, the gravitational drag from an energetic in-going particle (thick red arrow), acting on light out-going particles (thin, dashed arrows), is in the same upper-left direction by the same amount everywhere.\\ 
	$a2)$ dashed line: early Cauchy surface as seen by local observer; crossed line: later Cauchy surface.  In a black hole, however, diagrams $b$ apply:\\
	$b1)$ A global observer will interpret the drag differently. In region \(II\), all geodesics are in the opposite direction. The Cauchy surfaces, which we demand to stay complete, evolve as drawn as in $b2)$. See main text.}}
	\labell{Cauchy.fig}
	\end{caption}\end{quote}
\end{figure}

 Yet still something seemed to be not right. There is an apparent problem, as illustrated in Fig.~\ref{Cauchy.fig}. A particle arrives at the future event horizon (thick red arrow in Fig. \ref{Cauchy.fig}$a1$), at some fairly early moment in time. Its momentum in our present coordinates has increased so much that the gravitational drag effect becomes sizeable. For a local observer, this drag is the same for all out-going particles. However, this picture cannot be used to understand the effect of the drag on the Cauchy data as seen by a distant observer. In our initial calculations, the drag was inserted as in Fig.~\ref{Cauchy.fig}$b1$: the drag is still assumed to be the same everywhere, but since the out-going particles in region \(II\) `move backwards in time', the picture looks incorrect for the local observer. In particular, one out-going particle in the figure, drawn with a green line, seems to follow a geodesic that would be impossible for the local observer. What is going on?

We claim that Fig.~\ref{Cauchy.fig}$a1$ is incorrect for the global observer; it has to be replaced by \ref{Cauchy.fig}$b1$. We see in Fig.~\ref{Cauchy.fig}$b2$ that, in region \(II\), the Cauchy surfaces move in the opposite direction. If, at the same time, data are transported from region \(I\) to region \(II\) or \emph{vice versa}; these will continue their way by being transported into the new time direction. Only if we strictly adhere to the new rule, the data on the Cauchy surfaces stay complete. It is this completeness that is essential in the stage of the calculation where we insist to safeguard unitarity of the evolution law.  We simply have to state that the global observer is looking at a process where the \emph{local} observer sees an in-going particle with high momentum \(p^-\) entering the future event horizon, while at the other side of the horizon, the momentum \(p^-\) switches sign (see red dot in Fig.~\ref{Cauchy.fig}$a1$ and $b1$). We can see that it switches sign: the arrow is directed towards the horizon in region \(I\) but away from the horizon in region \(II\). Apparently, a particle was \emph{created} in region \(I\) and simultaneously \emph{annihilated} in region \(II\). This can be checked by noting that the out-going particles in region \(II\) are dragged away from the horizon (this even holds for Fig.~\ref{Cauchy.fig}$a1$). Apparently, \emph{all} geodesics are to be interpreted with the opposite signs in region \(II\), when compared with what they are in region \(I\). They go in the opposite space-time direction.

Our conclusion may seem to be wrong if we ask the local observer, but in that case, we observe that the local observer is not using the Cauchy surfaces that we need to consider\fn{He does use the Cauchy \emph{data} correctly.}, which are those of Fig.~\ref{Cauchy.fig}$b2$ rather than $a2$. We reached our conclusion by observing that, comparison of global phenomena to phenomena seen by local observers, has to be done with more care when we glue together the physical worlds at two sides of a horizon. Reaching a global physical law that is free from unitarity conflicts is our primary goal.

When comparing local with global observers, the local and the global observer must agree about the data on the Cauchy surface. They also agree on how these data are shifted when a fast moving in-particle is added. An inwards shift in region \(I\) implies an outward shift in region \(II\).

Note that, in any case, the coordinates \(u^\pm\) change sign when we go from a point in region \(I\) to its antipode in region \(II\), and, since momenta are dual to positions, it is therefore inevitable that also the momenta \(p^\pm\) change signs. Both positions and momenta are expanded in spherical harmonics \(Y_{\ell,m}\) with odd \(\ell\) only.

This topic was expanded rather extensively in this section, mainly because it may be teaching us a lesson: not to arrive at ambiguous conclusions too quickly. In this author's mind, the older views on quantum black holes consist of pictures that cannot be complete: when someone arrives at the conclusion that the data will be `chaotic' without the possibility to disentangle the logic even at a local level, then one may have to reconsider what our primary demands are for a theory to be consistent and complete.

As for the physical interpretation of the sign switch, see more about this in section~\ref{meeting}.
The prescription we found\,\cite{GtHrecent1,GtHrecent2} is unique and explicit, even if more work may be needed to explore all conceivable complications and generalisations. We expand on these in section~\ref{conc}.

\newsecl{The trapped regions}{trapped}
	We arrive at a precise prescription as to how familiar domains of theoretical physics are to be employed to formulate the evolution laws of quantum black holes, whenever \(M_\BH\gg M_\Pl\). Indeed, we start with the four postulates at the beginning of section~\ref{intro}. In short, we postulate the applicability of perturbative quantum field theory in a curved background, during short time intervals, assuming the presence exclusively of below-Planckian energy particles. Because their energies are low, we may assume that the background itself is a solution of Einstein's equations without matter (or perhaps very small amounts of matter). For a black hole this means that we employ the Penrose diagrams for eternal black holes (possibly generalised to the Kerr or Kerr-Newman configurations). Without any hindrance of singularities, we adopt the identification of region \(II\) with the antipodes of region \(I\), so that we again get a single, purely Minkowskian, asymptotic region. Since in this space-time, points near the horizon approach their antipodes very closely, we have a situation that, without antipodal identification,  could be interpreted as the non-locality that was expected to result from the postulates; with the antipodes on the horizon identified, our physics is local. 
	
	The firewalls\,\cite{firewall} can be addressed using the method further explained in Refs.\,\cite{GtHrecent1,GtHrecent2}. If one insists, this also could be regarded as  a departure from the requirements phrased earlier, but it is not much more than a careful reconsideration of how horizons should be treated. 
		
	A striking feature of our picture is that the \emph{trapped regions}, regions \(III\) and \(IV\) in our Penrose diagrams\fn{Note, that region \(IV\) can also be seen as identified with region \(III\) of the antipodal points.}, are never entered. Previous formalisms used what was called `nice slices' as Cauchy surfaces, but these carried the data far inside the trapped domains. The problem is that, exactly because they are trapped, the data on such Cauchy surfaces cannot be recovered, so that information loss indeed became inevitable.  Our procedure was especially designed so that the Cauchy surfaces evolve as in Figure \ref{Cauchy.fig}$b2$. This is their path in the absence of gravitational back reaction, and the back reaction is phrased in such a way that they continue to behave like this. The back reaction causes the data to be shifted around from region \(I\) to region \(II\) and back, but never shifts into regions \(III\) and \(IV\).
	
	We never need to enter into the trapped regions by more than an infinitesimal distance; rather than using such infinitesimal domains for the boundary conditions at the origin by demanding continuity, we may not need much more than our unitarity conditions for the Cauchy data.
	
\newsecl{What happens when a particle meets its antipode?}{meeting}
	As was explained when discussing Figure \ref{Cauchy.fig}, we were forced to flip the sign of the momentum \(p^-\) of the energetic particle entering the future event horizon. One can immediately ascribe this to our desire to keep all data on the combined Cauchy surfaces intact when gravity acts. Either data move from \(I\) to \(II\) or the converse happens. Data are lost when \(p^->0\), and it is gained when \(p^-<0\). So when the particle entering in region \(I\) has positive \(p^-\), the one entering\fn{Note, that these together form a single particle, whose wave function ranges over regions \(I\) and \(II\), both in position space and in momentum space.} in region \(II\) has negative \(p^-\). In our calculations\,\cite{GtHdiagonal,GtHrecent1,GtHrecent2}, this was built in: the coordinates \(u^\pm\) in region \(II\) are minus the coordinates in region \(I\). Now since \(p^\pm\) are the canonically associated variables, it is inevitable that these, in turn, also change sign when going from region \(I\) to region \(II\). This was also why, in our expansions in spherical harmonics, both the \(u^\pm\) variables and the \(p^\pm\) variables are all exclusively composed of \(Y_{\ell,m}\) with odd \(\ell\). The gravitational back reaction links \(u^\pm_{\ell,m}\) directly to \(p^\pm_{\ell,m}\). All of these only contain odd \(\ell\) values.	
	
	A question frequently asked is: 
	
	\emph{How can a black hole ever gain or loose energy, if the momenta add up to zero?}.
	
\noindent The answer is simple: \(p^\pm\) are not the momenta or the energies as seen by the distant observer. In fact, they are not sharply defined; they are the Fourier transforms of the wave functions over the entire stretch \(-\infty<u^\mp<\infty\), so these particles are superpositions of the particles and the antipodes. The wave functions may be chosen to have their support only in positive values of \(u^\mp\) or only their negative values. In that case, the particle is present only at one hemisphere of the black hole, with a slightly blurred momentum.

What is conserved is indeed the energy of the in- and out-going waves. These are obtained by Fourier transforming the waves (when still close to the horizon) not in \(u^\pm\) space, but in the values taken by \(\log |u^\pm|\), either in region \(I\) or in region \(II\).

So creating a particle entering the horizon with positive \(p^-\), is the same thing as annihilating a particle from the antipodal region (or creating a hole, having negative \(p^-\)). In both regions \(I\) and \(II\), half of position space is hidden from view; only by combining the two halves, one can perform the Fourier transformation to find the common value of \(p^-\), which is positive at one side, and negative for the other. The \emph{energy} added to the system by this wave function is sharply defined to be \(\k\).

The soft --\emph{ i.e.} low energy -- particles affected by the gravitational drag, also flip direction, as we see in the Figure. When regarded as a particle in region \(I\), a geodesic is pointing upwards, in region \(II\) it is downwards.\fn{Unfortunately, we did not manage to keep the same sign definitions of \(u^-\) in all our papers, as sometimes one definition is more convenient than an other -- my apologies to the reader. At this level of our inquiries, such sign definitions can be made any way we like; it is the relative signs that have to be carefully addressed.} At  all odd values of \(\ell\), the spherical waves oscillate between positive and negative values. This means that, in the spherical wave representation, all modes contain matter being created at some spots on the horizon, and/or matter being annihilated at the antipodal points. To find what this means in terms of the states in Fock space for the Standard Model (which may include low-energy gravitons), one simply has to evaluate the energy-momentum tensor components \(T^{--}\) and \(T^{++}\) as distributions across the horizons. This is an elementary exercise.

The more difficult question is, how to invert this mapping: given either \(T^{--}\) or the displacement operator associated to that, how can we recover the Fock space element? This question has not yet been fully answered. Counting arguments suggest that such an inversion procedure may be well-defined depending on the high angular momentum cut-off, both for the Fock space particles and the spherically harmonic waves.	It may well be an interesting exercise to investigate this dual transformation; this has not (yet) been done.

Note that, in the spherical harmonic expansion, the functions \(u^\pm(\tht,\vv)\) describe  sheets (two-branes, in modern terminology), whose values at definite points \((\tht_i,\,\vv_i)\) are the positions of particles, both in the space- and in the time direction (since both \(u^+\) and \(u^-\) are specified.\fn{Since this includes the time coordinate, one can state that indeed we have a two-dimensional world sheet, which has much in common with the world sheets of strings.}
	
\newsecl{Conclusion}{conc}
The author is quite confident that this approach is on the right track. We observed a number of indications that our descriptions are coherent and consistent.

In other lines of investigation,  conclusions were arrived at that may have to be reconsidered.  All we need is a basis of quantum states to formulate our theory in, and by invoking antipodal identification, a very apt basis, in the form of spherical harmonic amplitudes, was discovered. In our treatment, these spherical waves all decouple from one another, so that `chaos' does not play a significant role. Neither does `entanglement'; we should beware of the use of such buzz words. To this author, `chaos' is synonymous to `lack of understanding' (unless a local law, or a law for very short time intervals, can still be formulated). `Entanglement' is a magic wand that is less powerful than sometime suggested in the popular literature.

Even AdS/QFT did not provide us with the correct answers. This should not be read as  a disqualification of AdS/QFT, but rather as a warning: AdS/QFT  should also  be applied with the utmost care, and foolproof results should not be expected. We suspect that, given more appropriate boundary conditions, AdS/QFT may well still work. Note, however, that our asymptotic space-time is Minkowskian, not anti-deSitter, and string theory did not enter our work beyond a footnote.

This might change in the future. Our work is still in a development stage, and much more work must be done. One issue is the counting of micro-states. At first sight it may seem that the number of \(Y_{\ell,m}\) harmonic modes is unbounded, as also Fock space for elementary particles, even if they occur on a curved background. We do note that a bound \(\ell_\max\) is expected for \(\ell\), Eq.~\eqn{ellmax}. All ingredients needed to count these states correctly are there. Using Fermi's Golden Rule (``the probabilities of the final states are expressed as a quantum amplitude squared, multiplied with the volume of phase space for the final states"), we can calculate the probabilities and the amplitudes, to find the volume of phase space, even in the absence of thermal equilibrium. 

We noted earlier that the effects of our displacement operators \(u^\pm(\W)\) on the Fock space elements of our perturbative QFT on the Schwarzschild background, should be calculated. In principle, it is straightforward, but it has not been done. Also, quantum black hole considerations were the first that lead us to contemplate `holography' effects in physics. Indeed, the dynamical component of a quantum black hole is the horizon (a two-dimensional surface), while the interior, in our treatment, is absent. We still need to put all these ingredients in a proper perspective.

The question how the elements of Fock space for some particle model can be characterised if the distributions \(p^\pm(\W)\) and/or \(u^\pm(\W)\) are given, is an important one. In connection with this, an important follow-up question can be raised:
\begin{quote}
	\emph{If our position or momentum distributions are all expressed in terms of spherical harmonics with odd \(\ell\), how should we represent Hawking radiation, knowing that the distribution is dominated by very low energy particles in an \(\ell=0\) state?}
\end{quote}
Indeed, on average, a Hawking particle has just enough energy to overcome a small angular momentum barrier; if \(\ell\ge 3\), its contribution dwindles, due to the emerging grey-body factor. Hawking particles typically come as \(S\), \(P\) or \(D\) states.

The answer to the question is that the angular momentum quanta \((\ell,\,m)\) for a quantum particle should not be confused with the mode numbers \((\ell,\,m)\) for the momentum distributions as seen by local observers near the horizon(s). A precise answer will be given when we derive more precise relations between Fock space elements and spherical wave modes of the momentum an position distributions. We do have a preliminary answer. Going from the momentum or position operators back to Fock space, is most easily done by starting from the definition \eqn{momposdistr} --  \eqn{commrels} given in Section \ref{intro}. We imagine some grid in \(\W\) -space. Since we agree that dominant contributions come from the lower \(\ell\) values, the grid does not have to be a very fine one.

The fact that all \(\ell\) are odd means that, at every \(\W\), we have \(u^\pm(\W)=-u^\pm(-\W)\). This means that, at every \(\W\), we have not two but just one value for \(u^+\) and one for \(u^-\). The ones that are positive refer to region \(I\) at the point \(\W\), the ones that are negative refer to \(-\W\). Only where \(u^\pm\) are positive, the particle can be seen at all. Thus, at every projective \(\W\), we have a particle either at \(\W\) or at \(-\W\) but not both. All these particles only have positive values for \(u^\pm\). We could now decide to expand these positive values in terms of new spherical harmonics. Since now they all have the same (positive) sign, the \(\ell=0\) contribution certainly cannot vanish.

In short, the question pertains only to a notational feature. It was important to consider it here, since it indicates that the procedures to obtain Fock space states out of the momentum distribution modes, have to be carefully considered.

This is one of our major remaining problems,  which may be seen as the last step for our procedure. We found how matter entering a black hole leaves its information in the form of a momentum density on the future horizon. We found how the black hole transfers this information to position and/or momentum operator distributions for out going particles on the past horizon. However, on every point of the projective sphere of the horizon, this prescribes the presence of exactly one particle, either at the point \(\W\) or at its antipode \(-\W\), while Fock space usually allows for more than one particle at a given transverse coordinate. In our result, a multi-particle state at a given \(\W\) will be indistinguishable from a single particle state. This degeneracy must be lifted.

This situation is very similar to what we have in string theory. There also, a single vertex insertion operator \(e^{ik\cdot x}\) on the string world sheet does not distinguish multi-particle states from single particle states. Only after integrating over the locations of the vertices on the string world sheet, the full Veneziano amplitude emerges. What will be needed is a similar procedure for our approach. Thus, the problem does not seem to be hopelessly prohibitive, but it is something we will have to consider in detail. 

Just as in string theory, we may expect that the outcome of such formalism will lead to constraints concerning the structure of the Standard Model. In string theories, it was Lorentz invariance that brought about the constraints: the Virasoro algebra had to close correctly. In our case, it will be more difficult to understand conceptually how to implement Lorentz invariance since we have a Schwarzschild background. We do consider the questions treated in this paper as important ones. By insisting that black holes preserve the unitarity of quantum theories of gravity, we may be able to predict allowed forms of Standard Model interactions\fn{Besides imposing a prohibition for additively conserved quantum numbers such as lepton number, there will also be constraints on \(CP\) violation: the antipodal identification links space-time with its \(CP\) reflection. Our preliminary conclusion from that is that, in the Standard Model, \(CP\) can only be spontaneously broken, not explicitly.}. \\[20pt]
\noindent{\large \textbf{Acknowledgement}}\\

The author acknowledges discussions with P.~Betzios, G.~Dvali,  A.~Franzen,  N.~Gaddam,  S.~Giddings, S.~Hawking, J.~Maldacena, S.~Mathur, S.~Mukhanov, O.~Papadoulaki, F.~Scardigli, L.~Susskind, and W.~Vleeshouwers among others.

\end{document}